\documentstyle[prl,twocolumn,aps,epsf]{revtex}
\begin{document}

\title{New type of pion interferometry formula}

\author{ Q. H. Zhang} 

\address{ Institut f\"ur Theoretische Physik, Universit\"at 
Regensburg, D-93040 Regensburg, Germany}

\date{\today}

\maketitle

\begin{abstract}

New type of pion interferometry formulas is derived, in which 
not only the correlator but also the pion multiplicity 
is taken consideration. 
 The method to get information about initial emission probability of 
the unsymmetrized Bosons is given.
\end{abstract}


Two-particle Bose-Einstein(BE) interferometry 
 as a method to obtain information on the space-time geometry 
and dynamics of high energy 
collisions has recently received intensive theoretical and 
experimental attention\cite{Review}. 
However the old, widely used, two-pion interferometry formula has the 
shortcoming in that it does not take into account of multipion 
BE symmetrization effects, which have aroused great 
interests among physicists in the field of high energy 
physics\cite{LL,Zajc,Pratt93,CGZ,SL94,SBA97,Pol3,ZC1,CZ1,Zhang,Zhang2,Prc,Schlei,Pol2,Pol1,Weiner,Z98,Bialas} and 
statistical physics\cite{BC}. 
In this letter, we derive a new multipion 
interferometry formula which is determined not only by the correlator,  
but also explicitly by the pion multiplicity distribution. The main results 
of the letter are summarized as follows:
(1) If the bosons are emitted independently\cite{Bialas} and only BE 
symmetrization effects are present, the standard two-particle and higher-order 
interferometry formulas are reproduced. (2) If the bosons are not 
emitted independently(i.e., they are correlated even without BE symmetrization),
then these initial correlation will lead to a new pion interferometry 
formula which depends on the initial emission probability 
of the unsymmetrized Boson explicitly. 
(3) One can get information about initial emission 
probability of the unsymmetrized Boson through the study 
of the two-pion interferometry.

The multipion pion state can be written down as
\begin{equation}
|\phi \rangle =\sum_{n=0}^{\infty} a_n |n\rangle,
\end{equation}
where $a_n$ is a parameter connected with the pion multiplicity 
distribution. $|n\rangle$ is the n-pion state which can be expressed as
\begin{equation}
|n \rangle =\frac{(\int d{\bf p} \int j(x) a^+({\bf p})\exp(ip\cdot x) )^n}
{n!}|0\rangle.
\end{equation}
Here $a^+({\bf p})$ is the pion-creation operator and $j(x)$ is the 
pion current which can be expressed as:
\begin{equation}
j(x)=\int d^4y d^4p j(y,p)\gamma(y)\exp(-ip(x-y)),
\end{equation} 
where  $j(y,p)$ is the probability amplitude of finding a 
pion with momentum $p$, emitted by the emitter at $y$.
$\gamma(y)=\exp(i\phi(y))$ is a random phase factor which has been 
extracted from $j(y,p)$.
All emitters are uncorrelated in coordinate space 
when assuming\cite{Zhang}:
\begin{equation}
\{\gamma^*(x)\gamma(y)\}=\delta^4(x-y).
\end{equation}
Here $\{ \cdot \cdot \cdot \}$ means phase average. 
Eq.(4) corresponds to the assumption that the source is 
totally chaotic. If the source is totally coherent we have 
\begin{equation}
\phi(x)=0,~~ \gamma(x)=1.
\end{equation} 

According to Eq.(1), the normalized pion multiplicity distribution can be expressed as
\begin{equation}
P(n)=
\frac{|a_n|^2\{\langle n|n\rangle \}}{\sum_{n=0}^{\infty} |a_n|^2 \{\langle n|n\rangle \}}
\end{equation}
and the $n$ pion inclusive distribution reads
\begin{eqnarray}
&&N_i({\bf p_1,\cdot \cdot \cdot, p_i})
=\frac{1}{\sigma}\frac{d\sigma}
{d{\bf p_1}d{\bf p_2}\cdot\cdot\cdot d{\bf p_i}}
\nonumber\\
&&=
\frac{
\{\langle \phi |a^+({\bf p_1})\cdot \cdot \cdot a^+({\bf p_i})a({\bf p_i})\cdot \cdot\cdot
a({\bf p_1})|\phi \rangle
\}
}
{\{\langle \phi |\phi \rangle \}}
\end{eqnarray}
with
\begin{equation}
\int \prod_{k=1}^{i} d{\bf p_k} N_i({\bf p_1,\cdot\cdot\cdot,p_i})
=\langle n(n-1)\cdot\cdot\cdot(n-i+1) \rangle.
\end{equation}
$N_i({\bf p_1,\cdot \cdot \cdot, p_i})$ 
can be interpreted as the probability of finding $i$ pions with momentum 
${\bf p_1,\cdot \cdot \cdot, p_i}$ in a {\it given event}.
Then the $i$ pion correlation function reads\cite{MV97,ZPH98}:
\begin{equation}
C_i({\bf p_1,\cdot \cdot \cdot,p_i})=\frac{N_i({\bf p_1,\cdot \cdot \cdot,p_i})}
{\prod_{j=1}^{i}N_1({\bf p_j})}.
\end{equation}
In the following we study first multipion BE correlation effects on 
pion multiplicity distribution. Then 
 we will derive a new type $n$ pion correlation formula which depends not 
only on the correlator but also 
 explicitly on the pion multiplicity distribution. Finally 
we will study multipion BE correlation effects 
on two pion interferometry.

If $a_n=1, n\in (0,\infty)$ in Eq.(1) and the source 
is totally coherent, then we have
\begin{equation}
\{\langle n|n\rangle\}=\frac{n_0^n}{n!}
\end{equation}
with
\begin{equation}
n_0=\int |\tilde{j}({\bf p})|^2 d{\bf p},~~\tilde{j}({\bf p})=\int j(x)\exp(ipx)d^4x.
\end{equation}
The normalized unsymmetrized pion multiplicity distribution for above 
source can be expressed as (using Eq.(6))
\begin{equation}
P(n)=\frac{n_0^n}{n!}exp(-n_0).
\end{equation}
If $a_n=n!^{1/4}, n\in (0,\infty)$ and the source is 
coherent 
then we have 
\begin{equation}
P(n)=(\sum_{n=0}^{\infty}\frac{n_0^n}{\sqrt{n!}})^{-1}\cdot 
\frac{n_0^n}{\sqrt{n!}}.
\end{equation}
If $a_n=n!^{-1/8}, n\in (0,\infty)$ and the source 
is coherent then we have  
\begin{equation}
P(n)=(\sum_{n=0}^{\infty}\frac{n_0^n}{{n!}^{1.25}})^{-1}\cdot
\frac{n_0^n}{{n!}^{1.25}}.
\end{equation}
In the following, the above three states (corresponding 
to different choices of $a_n$) will be denoted as D1,D2 and D3 respectively. 
Here D1 corresponds to the case where the  pions are emitted independently
while D2 and D3 correspond to the case in which the pions are not emitted 
independently, even without BE symmetrization. 
In the following we will study multipion BE correlation effects on 
pion multiplicity distribution. Using the notation $\omega(n)=\{<n|n>\}$ and 
assuming the source is totally chaotic, 
we find\cite{CGZ} 
\begin{equation}
\omega(n)=
\frac{1}{n}\sum_{i=1}^{n}\omega(n-i)\int d{\bf p}G_i({\bf p},{\bf p}),
\end{equation} 
with
\begin{equation}
G_i({\bf p,q})=\int \rho({\bf p},{\bf p_1})
d{\bf p_1}\rho({\bf p_1},{\bf p_2})\cdot\cdot\cdot d{\bf p_{i-1}}\rho({\bf p_{i-1}},{\bf q}).
\end{equation}
Where $\rho({\bf p,q})$ is the 
Fourier transformation of the source distribution $g(x,K)$,
\begin{equation}
\rho({\bf p,q})=\int g(x,\frac{p+q}{2})exp(i(p-q)x)d^4x.
\end{equation}
We assume the source distribution as 
\begin{eqnarray}
g({\bf r},t,{\bf p})&=&
n_0\cdot (\frac{1}{\pi R^2})^{3/2}exp(-\frac{{\bf r}^2}{R^2})\delta(t)
\nonumber\\
&&
(\frac{1}{2\pi \Delta^2})^{3/2} exp(-\frac{{\bf p}^2}{2 \Delta^2}).
\end{eqnarray} 
Here $n_0$ is a parameter.
Using this source distribution, we can study multipion BE correlations effects 
on the pion multiplicity distribution, pion spectrum distribution and two-pion 
interferometry since 
$G_i({\bf p.q})$ can be calculated by recurence relations\cite{Pratt93,CGZ} or 
analytically\cite{ZC1,CZ1}. 
Multipion BE correlation effects on the pion multiplicity distribution are 
shown in Fig.1. One clearly sees that BE symmetrization shift the 
pion multiplicity distributions of D1,D2 and D3 to the right side.
The unsymmetrized pion multiplicity distribution for D1,D2 and D3 
are also shown in Fig.1. 

\vskip -0.0cm
\begin{figure}[h]\epsfxsize=8cm
\centerline{\epsfbox{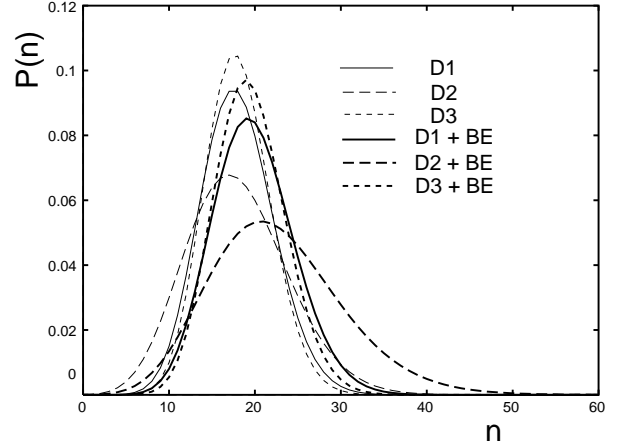}}
\caption{\it 
Multipion BE correlation effects on pion multiplicity distribution. 
The thin solid line, dashed line and dotted line correspond to 
the unsymmetrized pion multiplicity distribution of states D1,D2 and D3 respectively. 
The input values of $n_0$ for D1, D2 and D3 are $18$, $4.183$ and 
$37.38$ which ensure the mean unsymmetrized pion multiplicity 
 $\langle n \rangle_{D1}=\langle n \rangle_{D2}=\langle n\rangle_{D3}=18$.
The wider solid line, dashed line and dotted line correspond to the 
BE symmetrization effects on D1,D2 and D3 respectively.
The input values of $R$ and $\Delta$ are $5.3 fm$ and $0.18 Gev$.}
\end{figure}
According to Eq.(7), the one-pion inclusive distribution reads:
\begin{eqnarray}
N_1({\bf p_1})&=&
\frac{\{\langle \phi|a^+({\bf p_1})a({\bf p_1})|\phi\rangle\}}
{\{\langle \phi|\phi\rangle\}}
\nonumber\\
&=&\frac{\sum_{n=0}^{\infty}|a_n|^2\{\langle n|a^+({\bf p_1})a({\bf p_1})|n\rangle \}}
{\sum_{n=0}^{\infty}|a_n|^2\omega_n}.
\end{eqnarray}
Similar to Ref.\cite{CGZ}, one can obtain the following relation
\begin{equation}
\{\langle n|a^+({\bf p_1})a({\bf p_1})|n\rangle\}
=\sum_{i=1}^{n}\omega(n-i)G_i({\bf p_1,p_1}),
\end{equation}
the one-pion inclusive distribution can be expressed as:
\begin{equation}
N_1({\bf p})=\frac{\sum_{n=1}^{\infty}|a_n|^2\sum_{i=1}^{n}
\omega(n-i)G_i({\bf p,p})}{\sum_{n=0}^{\infty}|a_n|^2\omega(n)}.
\end{equation}
The two-pion inclusive distribution can be expressed as
\begin{eqnarray}
&&N_2({\bf p_1,p_2})=
\frac{ \{\langle \phi|a^+({\bf p_1})a^+({\bf p_2})a({\bf p_2})a({\bf p_1})|\phi\rangle\}}
{\{\langle \phi|\phi\rangle\}}
\nonumber\\
&&=\frac{\sum_{n=0}^{\infty}|a_n|^2\{
\langle n|a^+({\bf p_1})a^+({\bf p_2})a({\bf p_2})a({\bf p_1})|n\rangle \}}
{\sum_{n=0}^{\infty}|a_n|^2\omega_n}.
\end{eqnarray}
Using the relationship\cite{CGZ}
\begin{eqnarray}
&&\{\langle n|a^+({\bf p_1})a^+({\bf p_2})a({\bf p_2})a({\bf p_1})|n\rangle \}
=\sum_{i=2}^{n}\sum_{m=1}^{i-1}
[G_m({\bf p_1,p_1})
\nonumber\\
&&
G_{i-m}({\bf p_2,p_2})
+G_{m}({\bf p_1,p_2})G_{i-m}({\bf p_2,p_1})]\omega(n-i),
\end{eqnarray}
the two-pion inclusive distribution reads
\begin{eqnarray}
&&
N_2({\bf p_1,p_2})=
\frac{1}{\sum_{n=0}^{\infty}|a_n|^2\omega(n)}
\sum_{n=2}^{\infty}|a_n|^2\sum_{i=2}^{n}\sum_{m=1}^{i-1}
\nonumber\\
&&
[G_m({\bf p_1,p_1})
G_{i-m}({\bf p_2,p_2})+G_{m}({\bf p_1,p_2})
\nonumber\\
&&G_{i-m}({\bf p_2,p_1})]\omega(n-i)
.
\end{eqnarray}
It can be proved that 
the $k$-pion inclusive distribution can be expressed as
\begin{eqnarray}
&&N_k({\bf p_1,\cdot \cdot \cdot p_k})=
\frac{1}{\sum_{n=0}^{\infty}|a_n|^2\omega(n)}
\sum_{n=k}^{\infty}|a_n|^2\sum_{i=k}^{n}\sum_{m_1=1}^{i-(k-1)}
\nonumber\\
&&\sum_{m_2=1}^{i-m_1-(k-2)}\cdot\cdot\cdot 
\sum_{m_{k-1}=1}^{i-m_1-m_2\cdot\cdot\cdot-m_{k-2}-1}
\sum_{\sigma}
\nonumber\\
&&
G_{m_1}({\bf p}_1,{\bf p}_{\sigma(1)})
G_{m_2}({\bf p}_2,{\bf p}_{\sigma(2)})
\cdot \cdot\cdot
G_{m_{k-1}}({\bf p}_{k-1},{\bf p}_{\sigma(k-1)})
\nonumber\\
&&
G_{i-m_1\cdot\cdot\cdot-m_{k-1}}({\bf p}_{k},{\bf p}_{\sigma(k)})\omega(n-i)
\nonumber\\
&&
=
\frac{\sum_{n=i=m_1+m_2\cdot\cdot\cdot+m_k}^{\infty}|a_n|^2\omega(n-i)}{\sum_{n=0}^{\infty}|a_n|^2\omega(n)}
\sum_{\sigma}
\nonumber\\
&&
\sum_{m_1=1}^{\infty}G_{m_1}({\bf p_1,p_{\sigma(1)}})
\cdot\cdot\cdot\sum_{m_k=1}^{\infty}G_{m_k}({\bf p_k,p_{\sigma(k)}})
\nonumber\\
&&
=\sum_{n=i=m_1+m_2\cdot\cdot\cdot+m_k}^{\infty}P(n)\frac{\omega(n-i)}{\omega(n)}
\sum_{\sigma}
\nonumber\\
&&
\sum_{m_1=1}^{\infty}G_{m_1}({\bf p_1,p_{\sigma(1)}})
\cdot\cdot\cdot\sum_{m_k=1}^{\infty}G_{m_k}({\bf p_k,p_{\sigma(k)}})
\end{eqnarray}
Here $\sigma(k)$ denotes the $k$-th element of a permutations of the 
sequence $\{ 1,2,\cdot\cdot\cdot,k\}$, and the sum over $\sigma$ denotes 
the sum over all $k!$ permutations of this sequence.
It is interesting to notice that if $a_n=1 (P(n)=\omega(n)/\sum_n\omega(n))$ as 
assumed in Ref.\cite{Zhang2}, this new $k$ pion inclusive distribution are 
very similar to the old $k$ pion inclusive distribution which 
does not take account of  higher 
order BE correlation effects. This similarity warrants the 
validity of the formula used in earlier studies\cite{ELB,HZ97}. 
But in the general situation, pion interferometry formula depends not only on the correlator 
$\rho(i,j)$ but also explicitly on the pion multiplicity distribution $P(n)$ as shown in Eq.(25). 
Multipion BE correlation effects on 
the single pion inclusive distribution are shown in Fig.2. 
Here $P^{\langle n\rangle} ({\bf p})$ is
defined as 
\begin{equation}
P^{\langle n\rangle}(p)=\frac{N_1({\bf p})}{\langle n\rangle},
\langle n \rangle =\int N_1({\bf p})d{\bf p}.
\end{equation}
One clearly sees that multipion BE correlations cause pions to concentrate 
at the lower momenta region. Similar behavior was 
observed in Ref.\cite{ZC1}.  The above phenomena is caused by the nature 
of Bosons: pions would like to be in the same state. 
As the 
probability of finding a bigger $n$ pion state, $|n\rangle $, in D2
 is bigger than the probability of finding the same $n$ pion state in D3,
so the multipion 
BE correlation effects on D2 is stronger than the BE correlation effects 
on D3. 

\vskip -0.0cm
\begin{figure}[h]\epsfxsize=8cm
\centerline{\epsfbox{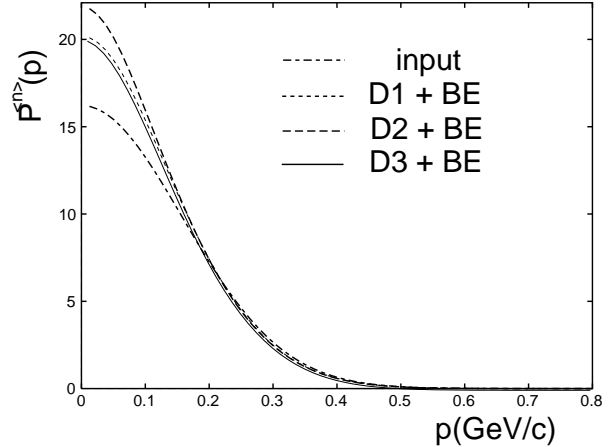}}
\caption{\it 
Multipion correlation effects on pion momentum distribution. The solid 
line, dashed line and dotted line correspond to multipion BE correlation effects 
on the pion spectrum distribution of pion state D3, D1 and D2 respectively. 
The mean pion multiplicity $\langle n \rangle=19$ for the above three cases. 
The dashed dotted line corresponds 
to the input momentum distribution $\int g(x,p) d^4x$. The input 
value of $R$ and $\Delta$ is $5 fm$ and $0.16 GeV$ respectively.}
\end{figure}

According to the definition of $n$ pion correlation function (Eq.(9)), 
one can easily write out $n$ pion interferometry formula by using 
Eq. (25).  
Multipion BE correlations effects on two-pion interferometry are shown in 
Fig.3. It is clear that one can tell the differences among those three 
different pion states 
through the 
studies of two-pion correlation function. For $D1$ state, the two-pion 
correlation function's intercept $C_2(q)_{q=0}$ with $q=|{\bf p_1}-{\bf p_2}|$ is $2$ 
and the value of two-pion correlation 
function is $1$ if the relative momentum $q$ becomes very large. While for D2 state,
one sees that $C_2(q)_{q=0}=2.19 > 2$ and $C_2(q)_{q\rightarrow \infty}=1.07 > 1$.
For D3 state, we have $C_2(q)_{q=0}=1.97 < 2$ and $C_2(q)_{q\rightarrow \infty}=0.987 < 1$.
 So one can 
tell the differences among the 
three type of multipion states by studying the behaviors of two-pion correlation function 
at smaller $q$ region and bigger $q$ region.
Because the resonance 
decay and the resolution power of the data, it is very difficult 
to determine the intercept of two-pion interferometry without further assumption.  
On the other hand, we can determine two-pion interferometry results at large q region 
without any assumption. 
As pion pairs at large q region is less affected by 
BE symmetrization so 
 the tail 
of two-pion interferometry is determined mainly by initial 
emission probability of unsymmetrized pions. According to Eqs.(12-14), 
one can determine $\frac{\langle n(n-1)\rangle}{\langle n\rangle^2}$ 
as 1.,1.07 and 0.986,for D1 ,D2 and D3 respectively.  Those values are 
consistent with the results obtained from two-pion interferometry.
Present heavy-ion experimental  
results indicates that $C_2(q)_{q\sim 300-400 MeV}$ is around $1.01 \sim 1.03$\cite{Comm1}. That 
means D1 is one of the best model (pion state)  to be 
used to analyze experimental results. 
\vskip -0.0cm
\begin{figure}[h]\epsfxsize=8cm
\centerline{\epsfbox{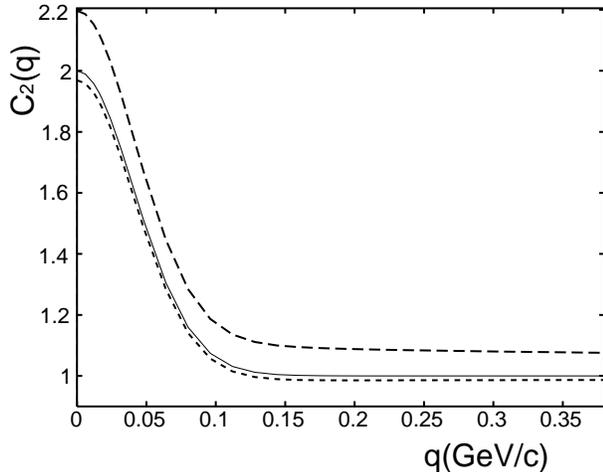}}
\caption{\it 
Multipion correlation effects on two-pion interferometry. The solid line,dashed line and 
dotted line corresponds to multipion BE correlation effects on two-pion 
interferometry for pion state D1,D2 and D3 respectively. The mean-pion 
multiplicity for three cases are $17.7$. The input value of 
 $R$ and $\Delta$ are $5 fm$ and $0.16 GeV$. } 
\end{figure}

Conclusion: In this Letter, multipion BE correlation effects on pion 
spectrum distribution and pion multiplicity distribution are studied.  
Due to multipion BE correlation effects, we derive a new type $n$ pion 
correlation function which is determined not only by 
correlator but also explicitly by the pion multiplicity distribution. 
This new pion interferometry formula is sensitive to the multiplicity 
distribution as well as to space structure of the source 
with and without BE symmetrization. 
We believe that one can get the information 
about pion states by pion interferometry method. 

\acknowledgements
I thanks Drs. U. Heinz, J.G. Cramer, T.A. Trainor, Y. Pang,
T. Cs\"org\H o, D. Mi\'skowiec, P. Scotto and Urs. Wiedemann 
for helpful discussions and communications. 
Q.H.Z. also would like to express his thanks to Dr. D. Storm 
for his hospitality during his visit at Nuclear Physics Laboratory.
The author owe thanks to Drs. J. Zim\'anyi, S. Pratt, H. Egger, B. Buschbeck,
Y. Sinyukov and R. Lednicky for discussions 
during CF'98 conference in Hungary. 
This work was supported by the Alexander 
von Humboldt Foundation. 
%

\begin{center}
{\bf Figure Captions}
\end{center}

\end{document}